\documentclass[12pt,oneside,american]{book}

\usepackage[T1]{fontenc}
\usepackage[latin9]{inputenc}
\usepackage[letterpaper]{geometry}
\usepackage{fancyhdr}
\usepackage{array}
\usepackage{float}
\usepackage{multirow}
\usepackage{amsmath}
\usepackage{amssymb}
\usepackage{mathdots}
\usepackage{setspace}
\PassOptionsToPackage{version=3}{mhchem}
\usepackage{mhchem}
\usepackage{esint}
\makeatletter

\providecommand{\tabularnewline}{\\}

\date{}
\usepackage{times}

\usepackage{amsthm}\usepackage{graphics}

\numberwithin{equation}{chapter}
\usepackage{babel}
\usepackage{titlesec} 

\titleformat{\chapter}{\centering\LARGE\bfseries}{\chaptertitlename\ \thechapter}{0.7em}{}
\titlespacing*{\chapter}{0pt}{0pt}{50pt}

\titleformat{\section}{\large\bfseries}{\thesection}{0.7em}{}
\titlespacing*{\section}{0pt}{12pt}{0pt}

\usepackage{indentfirst}
\usepackage{mathtools}

\makeatother

\usepackage{babel}
\begin{document}
\noindent \begin{center}
\textbf{\Large{}An MCMC Algorithm for Estimating the Q-matrix in a
Bayesian Framework}
\par\end{center}{\Large \par}

\bigskip{}
\noindent \begin{center}
Mengta Chung$^{1}$ and Matthew S. Johnson$^{2}$
\par\end{center}

\bigskip{}
\noindent \begin{center}
\textbf{\Large{}Abstract}
\par\end{center}{\Large \par}

\noindent The purpose of this research is to develop an MCMC algorithm
for estimating the Q-matrix. Based on the DINA model, the algorithm
starts with estimating correlated attributes. Using a saturated model
and a binary decimal conversion, the algorithm transforms possible
attribute patterns to a Multinomial distribution. Along with the likelihood
of an attribute pattern, a Dirichlet distribution, constructed using
Gamma distributions, is used as the prior to sample from the posterior.
Correlated attributes of examinees are generated using inverse transform
sampling. Closed form posteriors for sampling guess and slip parameters
are found. A distribution for sampling the Q-matrix is derived. A
relabeling algorithm that accounts for potential label switching is
presented. A method for simulating data with correlated attributes
for the DINA model is offered. Three simulation studies are conducted
to evaluate the performance of the algorithm. An empirical study using
the ECPE data is performed. The algorithm is implemented using customized
R codes.\textbf{\vspace{10pt}}

\noindent \textbf{Keywords} 

\noindent Q-matrix, DINA, CDM, Bayesian, MCMC
\noindent \begin{center}
\textemdash \textemdash \textemdash \textemdash \textemdash \textemdash \textemdash \textemdash \textemdash \textemdash \textemdash \textemdash \textemdash \textemdash \textemdash \textemdash \textemdash \textemdash \textemdash \textemdash \textemdash \textemdash \textemdash \textemdash \textemdash \textemdash \textemdash \textemdash \textemdash \textemdash \textemdash \textemdash \textemdash \textemdash \textemdash \textemdash \textemdash \textemdash \textemdash \textemdash \textemdash \textemdash \textemdash \textemdash \textemdash \textemdash \textemdash \textemdash \textemdash \textemdash \textemdash \textemdash \textemdash \textemdash \textemdash \textemdash \textemdash \textemdash \textemdash \textemdash \textemdash \textemdash \textemdash \textemdash \textemdash \textemdash \textemdash \textemdash \textemdash \textemdash \textemdash \textemdash \textemdash \textemdash \textemdash \textemdash \textemdash \textemdash \textemdash \textemdash \textemdash \textemdash \textemdash \textemdash \textemdash \textemdash \textemdash \textemdash \textemdash \textemdash \textemdash \textemdash \textemdash \textemdash \textemdash \textemdash \textemdash \textemdash \textemdash \textemdash \textemdash \textemdash \textemdash \textemdash \textemdash \textemdash \textemdash \textemdash \textemdash \textemdash \textemdash \textemdash \textemdash \textemdash \textemdash \textemdash \textemdash{}
\par\end{center}

\begin{singlespace}
\noindent This paper is based on the first author's doctoral dissertation:
\vspace{-30pt}
\end{singlespace}
\begin{description}
\begin{singlespace}
\item [{\textmd{Chung,}}] M. (2014). Estimating the Q-matrix for cognitive
diagnosis models in a Bayesian framework, Unpublished doctoral thesis,
Columbia University\vspace{-10pt}
\end{singlespace}
\end{description}
\begin{singlespace}
\noindent $^{1}$ Taipei Veterans General Hospital, mc3128@columbia.edu

\noindent $^{2}$ Columbia University, johnson@tc.columbia.edu
\end{singlespace}

\newpage{}
\noindent \begin{center}
\textbf{\Large{}Introduction}
\par\end{center}{\Large \par}

\noindent Cognitive diagnostic assessment (CDA) is a new framework
that aims to evaluate whether an examinee has mastered or possessed
a particular cognitive skill called an \textit{attribute} (Leighton
\& Gierl, 2007).  The last 20 years have seen the development of a
few cognitive diagnosis models (CDMs), such as the deterministic input,
noisy ``and'' gate (DINA) model (Junker \& Sijtsma, 2001), the noisy
input, deterministic ``and'' gate (NIDA) model (Maris, 1999), and
the reparameterized unified model (RUM) (DiBello, Stout, \& Roussos,
1995; Hartz, 2002). The core element of these models is the Q-matrix
(Tatsuoka, 1983), which is a binary matrix that establishes item-to-attribute
mapping in an exam. 

Traditionally the Q-matrix is fixed and designed by domain experts.
This could raise some issues. While some of the exams are written
with the purpose of being CDAs, their Q-matrices are not specified
during exam development and therefore have to be assigned after the
fact. Even when the Q-matrix is specified during the stage of exam
development, there are concerns that domain experts might neglect
some attributes or have different opinions. Therefore, it is of practical
importance to develop an automated method that offers a more objective
means of getting the Q-matrix. Despite the need of an automated Q-matrix
searching method, related research is still limited. The objective
of this research is to develop an MCMC algorithm for estimating the
Q-matrix in a Bayesian framework. Explicitly, we assume that the Q-matrix
is unknown and attempt to extract the entire Q-matrix from data.

A few studies that address the issue of Q-matrix have emerged (e.g.,
Barns, 2003; Winters, 2006; Templin \& Henson, 2006; Henson \& Templin,
2007; de la Torre, 2008; Desmarais, 2012; DeCarlo, 2012; Chiu, 2013;
Liu, Xu, \& Ying, 2012; Chen, Liu, Xu, \& Ying, 2015; Xu \& Desmarais,
2016). In particular, Templin and Henson (2006) advance a Bayesian
procedure to verify some uncertain Q-matrix entries for the DINA model.
In their procedure, uncertain Q-matrix entries in terms of subjective
probabilities are specified first, and posterior probabilities of
Q-matrix entries are the likelihood of an attribute required for a
successful response to an item. Subsequently, DeCarlo (2012) applies
the same Bayesian procedure to different Q-matrix conditions. However,
unlike Templin and Henson (2006), DeCarlo (2012) indicates that the
recovery rate is not always 100\% and the recovery is poor under the
situation of a complete uncertainty about an attribute. Nevertheless,
DeCarlo (2012) concludes that the Bayesian approach is in general
helpful to determine which attributes should be included or excluded
for each item. Extending Templin and Henson (2006) and DeCarlo (2012)
in an exploratory manner, we advance an MCMC algorithm for estimating
the whole Q-matrix.

A few other Q-matrix refinement and searching methods are also based
on the DINA model, such as de la Torre (2008), Liu, Xu, and Ying (2012),
Chen, Liu, Xu, and Ying (2015), and Xu and Desmarais (2016). The DINA
model, in which an examinee is viewed as either having or not having
a particular attribute, is parsimonious and easy to interpret. Whether
examinee $i$ possesses attribute $k$ is typically denoted as $\alpha_{ik}$,
a dichotomous latent response variable with values of $0$ or $1$
indicating absence or presence of a skill, respectively. The DINA
model is conjunctive. That is, in order to correctly answer item $j$,
examinee $i$ must possesses all the required attributes. Whether
examinee $i$ ($i=1,\cdots I$) is able to correctly answer item $j$
($j=1,\cdots,J)$ is defined by another latent response variable $\eta_{ij}$,

\begin{equation}\eta_{ij}=\prod\limits _{k=1}^{K}\alpha_{ik}^{q_{jk}}.\tag{1}\end{equation}The
latent response variable $\eta_{ij}$ is related to observed item
performance $X_{ij}$ according to the guess parameter, 
\[
g_{j}=P(X_{ij}=1|\eta_{ij}=0),
\]
and the slip parameter, 
\[
s_{j}=P(X_{ij}=0|\eta_{ij}=1).
\]

In other words, $g_{j}$ represents the probability of $X_{ij}=1$
when at least one required attribute is lacking, and $s_{j}$ denotes
the probability of $X_{ij}=0$ when all required attributes are present.
$1-s_{j}$ indicates the probability of a correct response for an
examinee classified as having all required skills. The item response
function (IRF) for item $j$ is \begin{equation}\ P(X_{ij}=1|\boldsymbol{\alpha}_{i})=(1-s_{j})^{\eta_{ij}}g_{j}^{1-\eta_{ij}}, \tag{2}\end{equation}and,
when local independence and independence among examinees are assumed,
the joint likelihood function for all responses is expressed as
\[
P(X_{ij}=x_{ij}|\boldsymbol{\alpha}_{i})=\prod\limits _{i=1}^{I}\prod\limits _{j=1}^{J}\biggl((1-s_{j})^{x_{ij}}s_{j}^{1-x_{ij}}\biggr)^{\eta_{ij}}\biggl(g_{j}^{x_{ij}}(1-g_{j})^{1-x_{ij}}\biggr)^{1-\eta_{ij}}.
\]

It should be noted that the monotonicity constraint, $1-s_{j}>g_{j}$,
should be placed in the estimation in order to enhance the interpretability
of the DINA model. Junker and Sijtsma (2001) observe that the monotonicity
does not always hold for the DINA model if no constraint is imposed.
\noindent \begin{center}
\textbf{\Large{}Proposed MCMC Algorithm}
\par\end{center}{\Large \par}

\noindent The setting for the estimation is comprised of item responses
from $I$ examinees to $J$ items that measure $K$ attributes. In
order to estimate the $J$ by $K$ Q-matrix, the following steps are
performed sequentially at iteration $t$, $t=1,\ldots,T$. The following
algorithm is implemented in base R (R Development Core Team, 2017). 

\noindent \textbf{Step 1: Binary Decimal Conversion}

With $K$ attributes, there are a total of $2^{K}$ \textit{possible}
attribute patterns for examinee $i$. Let $2^{K}=M$, and let the
matrix, $\boldsymbol{x}_{M\times K}=(x_{mk})_{M\times K}$, be the
binary matrix of \textit{possible} attribute patterns. Each of the
$M$ rows in $\boldsymbol{x}$ is a binary number that represents
a possible attribute pattern, which is converted to a decimal number
by $(b_{n}b_{n-1}\cdots b_{0})_{2}=b_{n}(2)^{n}+b_{n-1}(2)^{n-1}+\cdots+b_{0}(2)^{0},$
where $(b_{n}b_{n-1}\cdots b_{0})_{2}$ denotes a binary number.

After the conversion, these $M$ possible attribute patterns become
a Multinomial distribution. To estimate correlated attributes, a saturated
Multinomial model is used that assumes no restrictions on the probabilities
of the attribute patterns (see Maris, 1999). Assuming a Dirichlet
prior $\boldsymbol{\theta}$, the hierarchical model for estimating
attributes is\textbf{\vspace{-10pt}} 
\begin{align*}
\boldsymbol{x}|\boldsymbol{\theta} & \sim M\!ultinomial(M,\boldsymbol{\theta}),\\
\boldsymbol{\theta} & \sim Dirichlet(a_{1},a_{2},\ldots,a_{M}).
\end{align*}

\noindent \textbf{Step2: Updating Probability of Attribute Pattern}

Let $\boldsymbol{y}$ and $\boldsymbol{q}$ be the data and the Q-matrix.
Because the conjugate prior for a Multinomial distribution is a Dirichlet
distribution, the posterior $p(\boldsymbol{\theta}|\boldsymbol{x})\propto p(\boldsymbol{x}|\boldsymbol{\theta})p(\boldsymbol{\theta})$
is also a Dirichlet distribution. Therefore, use $Dirichlet(1,1,\ldots,1)$
as the prior, and the conditional posterior is distributed as $\mathit{Dirichlet}(1+y_{1},1+y_{2},\ldots,1+y_{M}),$
where $y_{\ell}$ $(\ell=1,\ldots,M)$ is the number of examinees
possessing the $\ell^{th}$ attribute pattern. As no function in base
R can be used to sample from the Dirichlet distribution, Gamma distributions
are used to construct the Dirichlet distribution. Suppose that $w_{1},\ldots,w_{M}$
are distributed as $\mathit{Gamma}(a_{1},1),\ldots,\mathit{Gamma}(a_{M},1)$,
and let $\tau=w_{1}+\cdots+w_{M}$. Then $(w_{1}/\tau,w_{2}/\tau,\ldots,w_{M}/\tau)$
is distributed as $\mathit{Dirichlet}(a_{1},a_{2},\ldots,a_{M})$. 

For each of the $M$ possible attribute patterns, we calculate the
total number of examinees $(y_{1},y_{2},\ldots,y_{M})$ falling into
an attribute pattern, and then sample from $\mathit{Gamma}(1+y_{1},1)=w'_{1},Gamma(1+y_{2},1)=w'_{2},\ldots,Gamma(1+y_{M},1)=w'_{M}$.
Let $\tau'=w'_{1}+w'_{2}+\cdots+w'_{M}$, and we can get the posterior
distribution $p(\boldsymbol{\theta}|\boldsymbol{x})\propto p(\boldsymbol{x}|\boldsymbol{\theta})p(\boldsymbol{\theta})=(w'_{1}/\tau',w'_{2}/\tau',\ldots,w'_{M}/\tau')$.
This posterior $p(\boldsymbol{\theta}|\boldsymbol{x})$ is used as
the prior $p(\boldsymbol{\theta})$ in the upper stage of the hierarchical
model. With the updated prior and the likelihood of each possible
attribute pattern, we obtain the full conditional posterior, $p(\boldsymbol{\theta}|\boldsymbol{y})\propto p(\boldsymbol{y}|\boldsymbol{\theta})p(\boldsymbol{\theta})=p(\boldsymbol{y}|\boldsymbol{\theta})(w'_{1}/\tau',w'_{2}/\tau',\ldots,w'_{M}/\tau')$.

\noindent \textbf{Step 3: Updating Attribute}

The full conditional posterior distribution is sampled using the discrete
version of inverse transform sampling. Let the posterior $(p_{1},p_{2},\text{\dots},p_{M})$
be the PMF of the $M$ possible attribute patterns. The CDF is computed
by adding up the probabilities for the $M$ points of the distribution.
To sample from this discrete distribution, we partition $(0,1)$ into
$M$ subintervals $(0,p_{1})$, $(p_{1},p_{1}+p_{2})$, \ldots{},
$(\sum\limits _{m=0}^{M}p_{m-1},\sum\limits _{m=0}^{M}p_{m})$, and
then generate a value $u$ from $\mathit{Uniform}(0,1)$. 

Updating the attribute state of examinee $i$ is achieved by checking
which subinterval the value $u$ falls into. This subinterval number
(a decimal number) is then converted to its corresponding binary number
(see step 1), which represents the attribute state of examinee $i$.
After steps 1 to 3 are carried out, attribute states for all examinees,
denoted as $\boldsymbol{\alpha}$, are obtained for iteration $t$.
It is noteworthy that the first 3 steps can also be used to estimate
$\boldsymbol{\alpha}$ in the NIDA model and the RUM (see Chung \&
Johnson, 2017). 

\noindent \textbf{Step 4: Updating Guess and Slip Parameters}

In general, posterior distributions are not available in closed forms
and therefore are usually approximated by MCMC sampling. The DINA
model has distinctive features, and we derive closed forms of the
full conditional posteriors for guess and slip parameters as follows.

With the estimated attribute states from step 3, this step updates
$g_{j}$ and $s_{j}$. $Beta(1,1)$, which is equal to $\mathit{Uniform}(0,1)$,
is chosen as the prior for both $g_{j}$ and $s_{j}$. Because the
conjugate prior for a Binomial distribution is a Beta distribution,
the full conditional posteriors of the guess and slip parameters are
also Beta distributions. In the DINA model, for examinee $i$ answering
item $j$, \textit{guess} occurs when $\eta_{ij}=0$ but $y_{ij}=1$,
and \textit{slip} happens when $\eta_{ij}=1$ but $y_{ij}=0$. Consequently,
in estimating $g_{j}$, the total number of \textit{successes} is
$\sum\limits _{i=1}^{I}(1-\eta_{ij})y_{ij}$, and the total number
of \textit{failures} is $\sum\limits _{i=1}^{I}(1-\eta_{ij})(1-y_{ij})$.
As $g_{j}\sim Beta(1,1)$ and $s_{j}\sim Beta(1,1)$, the full conditional
posterior distribution for $g_{j}$ is \begin{equation}
g_{j}|s_{j},\boldsymbol{\alpha},\boldsymbol{y},\boldsymbol{q}\sim\mathit{Beta}\Biggl(1+\sum\limits _{i=1}^{I}(1-\eta_{ij})y_{ij},1+\sum\limits _{i=1}^{I}(1-\eta_{ij})(1-y_{ij})\Biggr). \tag{3}
\end{equation}In estimating $s_{j}$, the total number of \textit{successes} is
$\sum\limits _{i=1}^{I}\eta_{ij}(1-y_{ij})$, and the total number
of \textit{failures} is $\sum\limits _{i=1}^{I}\eta_{ij}y_{ij}$.
Therefore, the full conditional posterior distribution for $s_{j}$
is \begin{equation}
s_{j}|g_{j},\boldsymbol{\alpha},\boldsymbol{y},\boldsymbol{q}\sim\mathit{Beta}\Biggl(1+\sum\limits _{i=1}^{I}\eta_{ij}(1-y_{ij}),1+\sum\limits _{i=1}^{I}\eta_{ij}y_{ij}\Biggr). \tag{4}
\end{equation}

The monotonicity constraint indicates that the probability of answering
an item correctly is supposed to be higher for an examinee who possesses
all the required attributes than for one who lacks at least one attribute,
that is, $1-s_{j}>g_{j}$. To achieve monotonicity, we use inverse
transform sampling to sample from a truncated Beta distribution.\textbf{
}The $g_{j}$ and $s_{j}$ parameters are sampled from $\mathit{Uniform}(0,1-s_{j})$
and $\mathit{Uniform}(0,1-g_{j})$, and then inverted to Beta distributions.\textbf{ }

Of note is that along the way to estimate the Q-matrix, steps 1 to
4 can be employed to estimate $\boldsymbol{\alpha}$, $\boldsymbol{g}$
and $\boldsymbol{s}$ when the Q-matrix is known.

\noindent \textbf{Step 5: Updating the Q-matrix}

Let $\boldsymbol{q}$ be the estimated Q-matrix from iteration $t-1$.
With the updated $\boldsymbol{\alpha}$, $\boldsymbol{g}$ and $\boldsymbol{s}$
from previous steps, step 5 updates the Q-matrix. Similar to step
1, this step uses a saturated Multinomial model to cope with correlated
attributes. With $K$ attributes, there are $2^{K}$ possible Q-matrix
patterns for item $j$. Because an item has to measure at least one
attribute, the pattern with all $0$'s has to be excluded, thus leaving
only $2^{K}-1$ possible patterns. Let $2^{K}-1=H$, and let $\boldsymbol{\epsilon}_{H\times K}=(\varepsilon_{hk})_{H\times K}$
be the matrix of \textit{possible} Q-matrix patterns for item $j$.
Accordingly, $\boldsymbol{\epsilon}$ has $H$ rows, and each row
of $\boldsymbol{\epsilon}$ represents a possible Q-matrix pattern.
Convert each of the $H$ possible Q-matrix patterns to a decimal number
(see step 1), and these patterns are distributed as a Multinomial
distribution. In updating the Q-matrix for item $j$, the model is
\begin{align*}
\boldsymbol{\epsilon}|\boldsymbol{\phi} & \sim M\!ultinomial(H,\boldsymbol{\phi}),\\
\boldsymbol{\phi} & \sim p(\boldsymbol{\phi}).
\end{align*}

Unlike step 2 that adopts a Dirichlet prior to estimate $\boldsymbol{\alpha}$,
step 5 uses the following approach in order to observe the underlying
probability of each Q-matrix entry. Denote an entry in the Q-matrix
as $q_{jk}$. Let $p(q_{jk}=1)=\phi_{jk}$ and $p(q_{jk}=0)=1-\phi_{jk}$.
Because the conjugate prior for a Bernoulli distribution is a Beta
distribution, $\mathit{Beta}(1,1)$ is chosen as the prior, $\phi_{jk}\sim Beta(1,1)$.
Therefore, the conditional posterior for $\phi_{jk}$ is distributed
as $\mathit{Beta}(1+q_{jk},2-q_{jk}).$ It is anticipated that the
posterior mean is $2/3$ for $q_{jk}=1$ and $1/3$ for $q_{jk}=0$.

Let $\boldsymbol{\phi}_{H\times K}=(\boldsymbol{\phi}_{1},\ldots,\boldsymbol{\phi}_{H})$,
where each element in the vector is a row in $\boldsymbol{\phi}$.
That is, $\boldsymbol{\phi}_{1}=(\phi_{11},\phi_{12},\cdots,\phi_{1K})$
and $\boldsymbol{\phi}_{H}=(\phi_{H1},\phi_{H2},\cdots,\phi_{HK})$.
Therefore, the prior for sampling from possible Q-matrix patterns
of item $j$ is distributed as \vspace{-20pt}

\[
p(\boldsymbol{\phi})\sim\left(\prod\limits _{k=1}^{K}\phi_{1k}^{\varepsilon_{1k}}(1-\phi_{1k})^{1-\varepsilon_{1k}},\prod\limits _{k=1}^{K}\phi_{2k}^{\varepsilon_{2k}}(1-\phi_{2k})^{1-\varepsilon_{2k}},\cdots,\prod\limits _{k=1}^{K}\phi_{Hk}^{\varepsilon_{Hk}}(1-\phi_{Hk})^{1-\varepsilon_{Hk}}\right).
\]
Each element in $p(\boldsymbol{\phi})$ is the probability of a possible
Q-matrix pattern for item $j$. The full conditional posterior distribution
is $p(\boldsymbol{\phi}|\boldsymbol{y})\propto p(\boldsymbol{y}|\boldsymbol{\phi})p(\boldsymbol{\phi})$.
With the likelihood for item $j$ from each of the $H$ possible patterns
and the prior $p(\boldsymbol{\phi})$, the Q-matrix for item $j$
can be sampled from the full conditional posterior. This sampled decimal
number is then converted to a binary number (see step 1), which is
the Q-matrix estimate for item $j$. 

After the procedure is applied to every item, the whole Q-matrix for
iteration $t$ is derived. As the number of iterations is T, there
is a total of T estimated Q-matrices, which are stored in a 3-dimensional
array $\mathcal{A}_{J\times K\times T}$.

\noindent \textbf{Step 6: Relabeling Q-matrix Estimates}

One potential issue in Bayesian Q-matrix estimation is label switching,
which arises when columns of the Q-matrix of the Bayesian model are
switched multiple times on different iterations during one run of
an MCMC sampling. Since the label sampled is assigned at each step
of the sampling, the assignment of the particular label is unique
only up to the permutation group (Jasra, Holmes, \& Stephens, 2005).
Label switching can be perceived as \textit{column} switching in the
Q-matrix estimation. For example, the following two Q-matrices are
equivalent even though the first column and the third column are switched,
\vspace{-10pt}

\[
\left[\begin{array}{ccc}
0 & 0 & 1\\
0 & 1 & 0\\
0 & 1 & 1\\
1 & 0 & 1
\end{array}\right]\thinspace\thinspace\thinspace\thinspace\thinspace\left[\begin{array}{ccc}
1 & 0 & 0\\
0 & 1 & 0\\
1 & 1 & 0\\
1 & 0 & 1
\end{array}\right].
\]
This raises concerns in the estimation. If label switching happens
during a run of MCMC, posterior summaries will be biased and have
inflated variance, although the result may match after columns are
relabeled. As a consequence, simply calculating the mean of these
T estimated Q-matrices from T iterations without relabeling might
yield a misleading final Q-matrix estimate.

Erosheva and Curtis (2017) propose a relabeling algorithm to account
for label switching in Bayesian confirmatory factor analysis. The
essential concept of their procedure is to relabel the factors after
the fact. We adopt the same concept and relabel each of the T estimated
Q-matrices stored in $\mathcal{A}_{J\times K\times T}$ from step
5. The logic of our procedure is as the following. Let $\mathcal{C}_{J\times K}$
be the average of the T estimated Q-matrices stored in $\mathcal{A}_{J\times K\times T}$
, and use $\mathcal{C}_{J\times K}$ as the first arbitrary reference.
The Euclidean distance is calculated from each permutation of a Q-matrix
estimate to $\mathcal{C}$. The permutation with the shortest Euclidean
distance is the relabeled Q-matrix $\mathcal{A}_{r}^{(t)}$,\begin{equation}
\mathcal{A}_{r}^{(t)}=\underset{\mathcal{A}^{(1)},\cdots,\mathcal{A}^{(K!)}}{\mathrm{min}}\Bigl[d(\mathcal{A}^{(k)}-\mathcal{C})\Bigr],\thinspace t=1,\cdots,T.\tag{2}
\end{equation} 

After each of the T estimated Q-matrices in $\mathcal{A}_{r}^{(t)}$
is relabeled and stored as $\mathcal{A}^{'}$, the average of these
T relabeled Q-matrices in $\mathcal{A}^{'}$ is the new arbitrary
reference $\mathcal{C}^{'}$. Using equation (2), $\mathcal{A}^{'}$
is relabeled again with $\mathcal{C}^{'}$ as the arbitrary reference.
This subroutine is run recursively until $\mathcal{A}^{'}$ converges.
The final Q-matrix estimate is then derived by calculating the average
of the T estimated Q-matrices stored in $\mathcal{A}^{'}$.\textbf{\vspace{10pt}}

\noindent \textbf{Summary of the Algorithm}

The algorithm is summarized as follows. With the binary decimal conversion,
possible attribute patterns are transformed to a saturated Multinomial
distribution (step 1). Along with the likelihood of an attribute pattern,
a Dirichlet distribution is used as the prior to sample from the posterior.
The Dirichlet distribution is constructed using Gamma distributions
(step 2), and attributes of examinees are updated using inverse transform
sampling (step 3). Sequentially, guess and slip parameters are generated
by Gibbs sampling using expressions (3) and (4) (step 4). The Q-matrix
is generated using a saturated Multinomial model (step 5). The final
Q-matrix is obtained after the relabeling algorithm in accomplished
(step 6).
\noindent \begin{center}
\textbf{\Large{}Simulation Study}
\par\end{center}{\Large \par}

\noindent \textbf{Procedure for Simulating Data}

\textbf{Generating Correlated Attributes}. Simulated data sets were
generated using the following procedure. The first step is to generate
correlated attributes. Let $\boldsymbol{\vartheta}$ be the $N$ by
$K$ underlying probability matrix of $\boldsymbol{\alpha}$, and
let column $k$ of $\boldsymbol{\vartheta}$ be a vector $\boldsymbol{\vartheta}_{k}$,
$k=1,\ldots,K$. That is, $\boldsymbol{\vartheta}=(\boldsymbol{\vartheta}{}_{1},\ldots,\boldsymbol{\vartheta}_{K})$.
A copula is used to generate intercorrelated $\boldsymbol{\vartheta}$
(see Ross, 2006). The correlation coefficient for each pair of columns
in $\boldsymbol{\vartheta}$ takes a constant value $\rho$ , and
the correlation matrix $\boldsymbol{\varSigma}$ is expressed as
\[
\boldsymbol{\varSigma}=\left[\begin{array}{ccc}
1 &  & \rho\\
 & \ddots\\
\rho &  & 1
\end{array}\right],
\]
where the off-diagonal entries are $\rho$. Each entry in $\boldsymbol{\varSigma}$
corresponds to the correlation coefficient between two columns in
$\boldsymbol{\vartheta}$. Symmetric with all the eigenvalues positive,
$\boldsymbol{\varSigma}$ is a real symmetric positive-definite matrix
that can be decomposed as $\boldsymbol{\varSigma}=\boldsymbol{\mathcal{\mathrm{\mathcal{\nu}}}}^{\mathrm{T}}\boldsymbol{\mathcal{\mathrm{\mathcal{\nu}}}}$
using Choleski decomposition, where $\mathcal{\boldsymbol{\mathcal{\mathrm{\mathcal{\nu}}}}}$
is an upper triangular matrix. 

After $\boldsymbol{\mathcal{\nu}}$ is derived, create an $I\times K$
matrix $\boldsymbol{\tau}$, in which each entry is generated from
$\mathit{N}(0,1)$. $\boldsymbol{\mathcal{\mathrm{\mathcal{\tau}}}}$
is then transformed to $\boldsymbol{\gamma}$ by using $\boldsymbol{\gamma}=\boldsymbol{\mathcal{\tau\boldsymbol{\nu}}}$,
so that $\boldsymbol{\mathcal{\mathrm{\gamma}}}$ and $\boldsymbol{\mathrm{\varSigma}}$
will have the same correlation structure. Set $\Phi(\boldsymbol{\mathcal{\mathrm{\gamma}}})=\boldsymbol{\vartheta}$,
where $\Phi(\cdot)$ is the CDF of the standard normal distribution.
To generate $\boldsymbol{\alpha}$, researchers have been using one
of the following two ways. Chen, Liu, Xu, and Ying (2015) generate
$\boldsymbol{\alpha}$ by \begin{equation}
\alpha_{ik}=\begin{cases} 1 & \textrm{if }\vartheta_{ik}\geq0\\ 0 & \textrm{otherwise} \end{cases}, \tag{5}
\end{equation}and Chiu, Douglas, and Li (2009) and Liu, Xu, and Ying (2012) use
the following criteria, \begin{equation}
\alpha_{ik}=\begin{cases} 1 & \textrm{if }\vartheta_{ik}\geq\Phi^{-1}(\frac{k}{K+1})\\ 0 & \textrm{otherwise} \end{cases}. \tag{6}
\end{equation}

\textbf{Generating Item Responses}. For the DINA model, $\boldsymbol{\eta}$
is determined by equation (1). After setting the guess and slip parameters
for each item, we can calculate the probability of an examinee correctly
answering an item by equation (2). An $N\times J$ probability matrix
$\boldsymbol{y}$ is thus formed, wherein each of the elements represents
the probability of an examinee correctly answering an item. Inverse
transform sampling for two categories, $0$ and $1$, is used to generate
the data. Create another $N\times J$ probability matrix $\boldsymbol{c}$,
with each element generated from $\mathit{Uniform}(0,1)$. These two
$N\times J$ matrices are then compared. If the corresponding value
in $\boldsymbol{y}$ is greater then that in $\boldsymbol{c}$, then
set $y_{nj}$ to 1; if otherwise, set $y_{nj}$ to 0. The final altered
$\boldsymbol{y}$ is the simulated data. Simply put,
\[
y_{nj}=\begin{cases}
1 & \textrm{if }y_{nj}\geq c_{nj}\\
0 & \textrm{otherwise}
\end{cases}.
\]

\noindent \textbf{Measure of Accuracy}

For $M$ simulated data sets, let $\hat{\boldsymbol{q}}^{(m)}=(\hat{q}_{jk}^{(m)})_{J\times K}$
$(m=1,\ldots,M)$ be the estimated Q-matrix from $m^{th}$ data set,
and let $\boldsymbol{q}=(q_{jk})_{J\times K}$ represents the true
$\boldsymbol{q}$. To measure how well the algorithm recovers the
true $\boldsymbol{q}$, the recovery rate $\Delta_{q}$, confined
between $0$ and $1$, is defined as \begin{equation}
\Delta_{q}=\frac{1}{M}\sum\limits _{m=1}^{M}\biggl(1-\frac{\Bigl|\hat{\boldsymbol{q}}^{(m)}-\boldsymbol{q}\Bigr|}{JK}\biggr),\:m=1,2,\ldots,M, \tag{7}
\end{equation}where $|\cdot|$ is the absolute value. 

\noindent \textbf{Settings for Simulation}

Congdon (2005) indicates that using single long runs may be adequate
only for straightforward problems, and Gelman and Shirley (2011) suggest
simulating three or more parallel chains in general. As estimating
the Q-matrix is a complicated process, we simulated 3 chains with
different random initial values.

Geyer (1991) points out that the accuracy of calculated quantities
depends on the adequacy of the burn-in period, which however can never
be validated for certain. Gelman and Shirley (2011) recommend discarding
the first half of simulated sequences as burn-in periods and mix all
the simulations from the second halves of the chains together to summarize
the target distribution so that the issue of autocorrelation is reduced.
We followed the advice advocated by Gelman and Shirley (2011). Corresponding
R codes were run 200,000 iterations after 200,000 burn-in periods
for each of the 3 chains. 

For each of the following simulations, examinees in groups of 500,
1000 and 2000 were simulated with the correlation between each pair
of attributes set to 0.1, 0.3 and 0.5. A hundred data sets were simulated
for each combination of sample size and correlation. The following
simulations were performed on 20 different Mac Pro computers, each
of which equipped an 8-core Intel Xeon E5 processor and 32 GB memory.

\noindent \textbf{Simulation I}

The first simulation serves to see how the algorithm performs in a
simple condition. The Q-matrix for simulation I is exhibited on the
left side of Table 1. This artificial Q-matrix (Q-matrix I) is obtained
from Rupp and Templin (2008). Fifteen items measuring 4 attributes
comprise a Q-matrix manifesting a clear pattern, which is constructed
in such a way that each attribute appears alone from items 1 to 4,
in a pair from items 5 to 10 , in triplicate from items 11 to 14 and
in quadruplet on item 15.

This Q-matrix is balanced, as each attribute is measured by 12 items.
This Q-matrix is complete, containing at least one item devoted solely
to each attribute (see Chiu, Douglas, \& Li, 2009). On average, each
item measures 2.133 attributes. In generating the data for simulation
I, $\boldsymbol{\alpha}$ was determined using equation (5), which
suggested the same difficulty level for each attribute. Guess and
slip parameters were set to 0.2 for all items in generating data.

\noindent \textbf{Simulation II}

In reality, different attributes could have different levels of difficulty.
The purpose of simulation II is to see whether using more complicated
cutoff criteria in generating $\boldsymbol{\alpha}$ affects the recovery
of the Q-matrix. The second simulation also used Q-matrix I. The difference
between simulation I and simulation II was that $\boldsymbol{\alpha}$
was generated using equation (6) instead of equation (5). Assuming
each attribute has a different difficulty level, equation (6) is more
complicated than equation (5) that regards each attribute as having
the same difficulty level. Specifically, equation (6) implies that
attribute 5 is the most difficult while attribute 1 is the easiest.
Guess and slip parameters were also set to 0.2 for all items as in
simulation I.

\noindent \textbf{Simulation III}

In addition to using the more complicated equation (6) to generate
$\boldsymbol{\alpha}$, simulation III uses a more intricate Q-matrix.
On the right side of Table 1 is the contrived Q-matrix (Q-matrix II)
for the third simulation. This 15 by 5 Q-matrix is modified from the
Q-matrix offered by de la Torre (2009). We excluded the first half
of the original Q-matrix and retained the remaining 15 items (items
16 to 30) to make it imbalanced and incomplete. Q-matrix II was imbalanced
in that each attribute appeared a different number of times in each
item (6, 8, 8, 9, 9 times). Q-matrix II was incomplete, because it
did not include items that measure each attribute alone. Each item
measures at least 2 attributes. On average, each item measured 2.67
attributes. Like simulations I and II, simulation III set guess and
slip parameters to 0.2 for all items.

\begin{table}[h]
\noindent \begin{centering}
Table 1. Q-matrices for Simulations
\par\end{centering}
\centering%
\begin{tabular}{cccccccccccccc}
\hline 
\multicolumn{6}{c}{Q-matrix I} &  & \multicolumn{7}{c}{Q-matrix II}\tabularnewline
\cline{1-6} \cline{8-14} 
\multirow{2}{*}{Item} &  & \multicolumn{4}{c}{Attribute} &  & \multirow{2}{*}{Item} &  & \multicolumn{5}{c}{Attribute}\tabularnewline
\cline{3-6} \cline{10-14} 
 &  & 1 & 2 & 3 & 4 &  &  &  & 1 & 2 & 3 & 4 & 5\tabularnewline
\cline{1-1} \cline{3-6} \cline{8-8} \cline{10-14} 
1 &  & 1 & 0 & 0 & 0 &  & 1 &  & 0 & 1 & 0 & 1 & 0\tabularnewline
2 &  & 0 & 1 & 0 & 0 &  & 2 &  & 0 & 1 & 0 & 0 & 1\tabularnewline
3 &  & 0 & 0 & 1 & 0 &  & 3 &  & 0 & 0 & 1 & 1 & 0\tabularnewline
4 &  & 0 & 0 & 0 & 1 &  & 4 &  & 0 & 0 & 1 & 0 & 1\tabularnewline
5 &  & 1 & 1 & 0 & 0 &  & 5 &  & 0 & 0 & 0 & 1 & 1\tabularnewline
6 &  & 1 & 0 & 1 & 0 &  & 6 &  & 1 & 1 & 1 & 0 & 0\tabularnewline
7 &  & 1 & 0 & 0 & 1 &  & 7 &  & 1 & 1 & 0 & 1 & 0\tabularnewline
8 &  & 0 & 1 & 1 & 0 &  & 8 &  & 1 & 1 & 0 & 0 & 1\tabularnewline
9 &  & 0 & 1 & 0 & 1 &  & 9 &  & 1 & 0 & 1 & 1 & 0\tabularnewline
10 &  & 0 & 0 & 1 & 1 &  & 10 &  & 1 & 0 & 1 & 0 & 1\tabularnewline
11 &  & 1 & 1 & 1 & 0 &  & 11 &  & 1 & 0 & 0 & 1 & 1\tabularnewline
12 &  & 1 & 1 & 0 & 1 &  & 12 &  & 0 & 1 & 1 & 1 & 0\tabularnewline
13 &  & 1 & 0 & 1 & 1 &  & 13 &  & 0 & 1 & 1 & 0 & 1\tabularnewline
14 &  & 0 & 1 & 1 & 1 &  & 14 &  & 0 & 1 & 0 & 1 & 1\tabularnewline
15 &  & 1 & 1 & 1 & 1 &  & 15 &  & 0 & 0 & 1 & 1 & 1\tabularnewline
\hline 
\end{tabular}

\end{table}

\noindent \textbf{Results}

The recovery rate of each concoction before and after relabeling is
exhibited in Table 2. Note that if the improvement of recovery rate
was less than 0.001, we did not list the recovery rate before the
relabeling algorithm was applied.

The recovery rate for each combination in simulation I was above 0.990,
suggesting that this MCMC algorithm should be effective when the difficulty
of each attribute is the same and the Q-matrix is complete. No label
switching was found in simulation I even when the sample size was
as small as 500 and the correlation was as high as 0.5.

Compared with simulation I, simulation II had a lower recovery rate
ranging from 0.831 to 0.994. In general, when the sample size increases,
the recovery rate also increases; when the correlation increases,
the recovery rate decreases. Unlike simulation I, simulation II saw
label switching under some conditions. The biggest improvement in
recovery rate was 1.4\%, which was the result from a sample size of
500 with correlation 0.5.

For simulation III that used an incomplete and imbalanced Q-matrix,
results are shown on the right side of Table 2. It can be seen that
the recovery rate in simulation III, ranging from 0.822 to 0.843,
was the worst among the three simulations. Results show that the recovery
rate increases with sample size and decreases with attribute correlation.
Label switching was observed. Even though the trend was not very obvious,
label switching seemed to prone to occur when the sample size was
decreased and the correlation was increased. When the sample size
was 500 with correlation between each pair of attributes equal to
0.5, the recovery rate increased the by 6.2\% after the relabeling
algorithm, the highest increase of all the combinations. 

\begin{table}
\begin{onehalfspace}
\noindent \begin{centering}
Table 2. Recovery Rate
\par\end{centering}
\end{onehalfspace}
\begin{centering}
{\footnotesize{}\centering}{\small{}}%
\begin{tabular}{ccccccccccccc}
\hline 
 &  & \multicolumn{3}{c}{{\small{}Simulation I}} &  & \multicolumn{3}{c}{{\small{}Simulation II}} &  & \multicolumn{3}{c}{{\small{}Simulation III}}\tabularnewline
\cline{3-5} \cline{7-9} \cline{11-13} 
{\small{}Sample} &  & \multicolumn{3}{c}{{\small{}Correlation}} &  & \multicolumn{3}{c}{{\small{}Correlation}} &  & \multicolumn{3}{c}{{\small{}Correlation}}\tabularnewline
\cline{3-5} \cline{7-9} \cline{11-13} 
{\small{}Size} &  & {\small{}0.1} & {\small{}0.3} & {\small{}0.5} &  & {\small{}0.1} & {\small{}0.3} & {\small{}0.5} &  & {\small{}0.1} & {\small{}0.3} & {\small{}0.5}\tabularnewline
\cline{1-1} \cline{3-5} \cline{7-9} \cline{11-13} 
\noalign{\vskip\doublerulesep}
\multirow{2}{*}{{\small{}500}} &  & \multirow{2}{*}{{\small{}0.997}} & \multirow{2}{*}{{\small{}0.996}} & \multirow{2}{*}{{\small{}0.994}} &  & {\small{}(0.915)} & {\small{}(0.867)} & {\small{}(0.817)} &  & {\small{}(0.751)} & {\small{}(0.730)} & {\small{}(0.696)}\tabularnewline
 &  &  &  &  &  & {\small{}0.921} & {\small{}0.876} & {\small{}0.831} &  & {\small{}0.800} & {\small{}0.783} & {\small{}0.758}\tabularnewline
\multirow{2}{*}{{\small{}1000}} &  & \multirow{2}{*}{{\small{}1.000}} & \multirow{2}{*}{{\small{}0.998}} & \multirow{2}{*}{{\small{}0.996}} &  & \multirow{2}{*}{{\small{}0.963}} & {\small{}(0.928)} & {\small{}(0.883)} &  & {\small{}(0.838)} & {\small{}(0.816)} & {\small{}(0.769)}\tabularnewline
 &  &  &  &  &  &  & {\small{}0.932} & {\small{}0.888} &  & {\small{}0.846} & {\small{}0.825} & {\small{}0.781}\tabularnewline
\multirow{2}{*}{{\small{}2000}} &  & \multirow{2}{*}{{\small{}1.000}} & \multirow{2}{*}{{\small{}1.000}} & \multirow{2}{*}{{\small{}0.998}} &  & \multirow{2}{*}{{\small{}0.994}} & \multirow{2}{*}{{\small{}0.968}} & {\small{}(0.927)} &  & {\small{}(0.860)} & {\small{}(0.839)} & {\small{}(0.810)}\tabularnewline
 &  &  &  &  &  &  &  & {\small{}0.929} &  & {\small{}0.861} & {\small{}0.841} & {\small{}0.813}\tabularnewline
\hline 
\end{tabular}
\par\end{centering}{\small \par}
\noindent \begin{centering}
\textit{\small{}\vspace{0.5pt}
}
\par\end{centering}{\small \par}
\raggedright{}\textit{\small{}\quad{}Note: Numbers in parenthesis
is the recovery rate before relabeling}{\small \par}
\end{table}
\noindent \begin{center}
\textbf{\Large{}Empirical Study}
\par\end{center}{\Large \par}

\noindent \textbf{The ECPE Data} 

A standardized English as a foreign language examination, the Examination
for the Certificate of Proficiency in English (ECPE) is recognized
in several countries as official proof of advanced proficiency in
English (ECPE, 2015). Obtained from the CDM R package, the data consists
of responses of 2922 examinees to 28 multiple choice items that measure
3 attributes{\small{} }(morphosyntactic, cohensive, lexical) in the
grammar section of the ECPE. The data has been analyzed by Feng, Habing
and Huebner (2013), Templin and Hoffman (2013) and Templin and Bradshaw
(2014). It consists of the responses of 2,922 examinees to 28 multiple-choice
questions in the grammar section of the ECPE. We tentatively tried
to extract the Q-matrix from the ECPE data. In analyzing the data,
the current MCMC algorithm was run 400,000 iterations, in which the
first 200,000 were discarded as burn-in periods.

\noindent \textbf{Initial Values}

In estimating the Q-matrix for the ECPE data, we referred to the Q-matrix
(Table 3) obtained from Templin and Bradshaw (2014) that assumes 28
items measuring 3 attributes as the initial value to reflect our prior
knowledge. According to Templin and Bradshaw (2014), these 3 attributes
represent: (1) morphosyntactic rules, (2) cohensive rules, (3) lexical
rules. For other parameters, initial values were randomly assigned
as in the simulation studies.

\noindent \textbf{Results}

The estimated Q-matrix is given on the right side of Table 3. If the
Q-matrix suggested by Templin and Bradshaw (2014) is assumed to be
the true Q-matrix, 52 out of the 84 entries were correct in the estimation
when the cutoff was set to 0.5. The recovery rate of the estimated
Q-matrix was about 62\%. For attributes 1 to 3, the number of incorrect
estimates were respectively 5, 11 and 16. Among the 32 incorrect estimates,
12 entries had estimated values of less than 0.5 whereas the correct
values would have been 1's; 20 entries had estimated values above
0.5 whereas the correct entries would have been 0's. The Akaike information
criterion (AIC) is 85812.92 for the true Q-matrix and 85693.58 for
the estimated Q-matrix, suggesting that the estimated Q-matrix fits
the data better than the initial Q-matrix.

\begin{table}[H]
\begin{onehalfspace}
\noindent \begin{centering}
Table 3. Estimated Q-matrix for the ECPE Data 
\par\end{centering}
\end{onehalfspace}
\begin{centering}
{\footnotesize{}\centering}%
\begin{tabular}{ccccccccc}
\hline 
\noalign{\vskip3pt}
 &  & \multicolumn{3}{c}{Initial\textsuperscript{1}} &  & \multicolumn{3}{c}{Estimated}\tabularnewline
\cline{3-5} \cline{7-9} 
\noalign{\vskip3pt}
\multirow{1}{*}{Item} &  & 1 & 2 & 3 &  & {\small{}1} & 2 & 3\tabularnewline
\cline{1-1} \cline{3-5} \cline{7-9} 
\noalign{\vskip2pt}
{\small{}1} &  & 1 & 1 & 0 &  & {\small{}(1) 0.988} & {\small{}(1) 0.622} & {\small{}(1) 0.999}\tabularnewline[3pt]
\noalign{\vskip2pt}
{\small{}2} &  & 0 & 1 & 0 &  & {\small{}(1) 0.584} & {\small{}(1) 0.976} & {\small{}(0) 0.000}\tabularnewline[3pt]
\noalign{\vskip2pt}
{\small{}3} &  & 1 & 0 & 1 &  & {\small{}(1) 1.000} & {\small{}(1) 0.804} & {\small{}(1) 1.000}\tabularnewline[3pt]
\noalign{\vskip2pt}
{\small{}4} &  & 0 & 0 & 1 &  & {\small{}(0) 0.000} & {\small{}(1) 1.000} & {\small{}(0) 0.000}\tabularnewline[3pt]
\noalign{\vskip2pt}
{\small{}5} &  & 0 & 0 & 1 &  & {\small{}(0) 0.000} & {\small{}(1) 1.000} & {\small{}(0) 0.000}\tabularnewline[3pt]
\noalign{\vskip2pt}
{\small{}6} &  & 0 & 0 & 1 &  & {\small{}(0) 0.000} & {\small{}(1) 1.000} & {\small{}(0) 0.000}\tabularnewline[3pt]
\noalign{\vskip2pt}
{\small{}7} &  & 1 & 0 & 1 &  & {\small{}(1) 1.000} & {\small{}(0) 0.033} & {\small{}(0) 0.000}\tabularnewline[3pt]
\noalign{\vskip2pt}
{\small{}8} &  & 0 & 1 & 0 &  & {\small{}(0) 0.001} & {\small{}(1) 1.000} & {\small{}(0) 0.006}\tabularnewline[3pt]
\noalign{\vskip2pt}
{\small{}9} &  & 0 & 0 & 1 &  & {\small{}(0) 0.000} & {\small{}(0) 0.002} & {\small{}(1) 1.000}\tabularnewline[3pt]
\noalign{\vskip2pt}
{\small{}10} &  & 1 & 0 & 0 &  & {\small{}(1) 1.000} & {\small{}(0) 0.206} & {\small{}(1) 1.000}\tabularnewline[3pt]
\noalign{\vskip2pt}
{\small{}11} &  & 1 & 0 & 1 &  & {\small{}(0) 0.000} & {\small{}(0) 0.000} & {\small{}(1) 1.000}\tabularnewline[3pt]
\noalign{\vskip2pt}
{\small{}12} &  & 1 & 0 & 1 &  & {\small{}(1) 1.000} & {\small{}(0) 0.321} & {\small{}(1) 1.000}\tabularnewline[3pt]
\noalign{\vskip2pt}
{\small{}13} &  & 1 & 0 & 0 &  & {\small{}(1) 1.000} & {\small{}(0) 0.001} & {\small{}(0) 0.000}\tabularnewline[3pt]
\noalign{\vskip2pt}
{\small{}14} &  & 1 & 0 & 0 &  & {\small{}(1) 1.000} & {\small{}(1) 0.991} & {\small{}(1)0.999}\tabularnewline[3pt]
\noalign{\vskip2pt}
{\small{}15} &  & 0 & 0 & 1 &  & {\small{}(0) 0.000} & {\small{}(1) 1.000} & {\small{}(0) 0.000}\tabularnewline[3pt]
\noalign{\vskip2pt}
16 &  & 1 & 0 & 1 &  & {\small{}(1)}1.000 & {\small{}(1)} 0.673 & {\small{}(0) }0.000\tabularnewline[3pt]
\noalign{\vskip2pt}
17 &  & 0 & 1 & 1 &  & {\small{}(0) }0.055 & {\small{}(1)} 0.959 & {\small{}(0) }0.001\tabularnewline[3pt]
\noalign{\vskip2pt}
18 &  & 0 & 0 & 1 &  & {\small{}(0) }0.000 & {\small{}(1)} 1.000 & {\small{}(0) }0.000\tabularnewline[3pt]
\noalign{\vskip2pt}
19 &  & 0 & 0 & 1 &  & {\small{}(0) }0.000 & {\small{}(0) }0.000 & {\small{}(1)} 1.000\tabularnewline[3pt]
\noalign{\vskip2pt}
20 &  & 1 & 0 & 1 &  & {\small{}(1)} 1.000 & {\small{}(0) }0.000 & {\small{}(1)} 1.000\tabularnewline[3pt]
\noalign{\vskip2pt}
21 &  & 1 & 0 & 1 &  & {\small{}(0) }0.000 & {\small{}(1)} 1.000 & {\small{}(1)} 1.000\tabularnewline[3pt]
\noalign{\vskip2pt}
22 &  & 0 & 0 & 1 &  & {\small{}(1)} 1.000 & {\small{}(0) }0.003 & {\small{}(0) }0.000\tabularnewline[3pt]
\noalign{\vskip2pt}
23 &  & 0 & 1 & 0 &  & {\small{}(0) }0.000 & {\small{}(1)} 1.000 & {\small{}(0) }0.000\tabularnewline[3pt]
\noalign{\vskip2pt}
24 &  & 0 & 1 & 0 &  & {\small{}(0) }0.000 & {\small{}(1)} 0.999 & {\small{}(1)}1.000\tabularnewline[3pt]
\noalign{\vskip2pt}
25 &  & 1 & 0 & 0 &  & {\small{}(1)} 1.000 & {\small{}(1)} 0.920 & {\small{}(1)}0.989\tabularnewline[3pt]
\noalign{\vskip2pt}
26 &  & 0 & 0 & 1 &  & {\small{}(1)} 1.000 & {\small{}(1)} 0.694 & {\small{}(0) }0.002\tabularnewline[3pt]
\noalign{\vskip2pt}
27 &  & 1 & 0 & 0 &  & {\small{}(1)} 1.000 & {\small{}(0) }0.082 & {\small{}(1)} 1.000\tabularnewline[3pt]
\noalign{\vskip2pt}
28 &  & 0 & 0 & 1 &  & {\small{}(0) }0.000 & {\small{}(0) }0.000 & {\small{}(1)} 1.000\tabularnewline[3pt]
\hline 
\end{tabular}
\par\end{centering}
\noindent \raggedright{}\textit{\small{}Note: Initial}\textsuperscript{1}\textit{\small{}
is the Q-matrix obtained from }\textit{Templin and Bradshaw (2014)}\textit{\small{};
Numbers in parenthesis are the estimates rounded to the nearest whole
number.}{\small \par}
\end{table}
\noindent \begin{center}
\textbf{\Large{}Discussion}
\par\end{center}{\Large \par}

\noindent We advance an MCMC algorithm for estimating the Q-matrix
in a Bayesian framework. This automated Q-matrix searching procedure
is based on the DINA model. A prominent discovery is that closed form
posterior distributions for generating guess and slip parameters are
found. This not only conveys a delicate statistical characteristic
of the DINA model but also facilitates the speed of the algorithm. 

In sampling attributes and the Q-matrix, 2-stage hierarchical Multinomial
models are used. Saturated Multinomial models appear to be useful
in coping with correlated attributes, and the relabeling procedure
to account for label switching seems to improve the recovery rate.
Our findings from the simulation studies indicate that sample size,
degree of correlation, difficulty of attributes and structure of Q-matrix
all influence the recovery rate. In addition, label switching indeed
occurs in the estimation; however it is not as severe as we at first
supposed. 

Some limitations of this research and recommendations for future work
are the following. First, this research was not entirely exploratory
as we assumed that the number of attributes was known. Calculating
log-likelihood might be able to reveal how the estimated Q-matrix
with any given number of attributes fits the data. Second, the correlation
for each pair of attributes is fixed for each of the simulations.
More complicated correlation structures are needed to examine how
they affect the Q-matrix recovery. Applying Choleski decomposition,
along with Dirichlet priors, to estimating the Q-matrix might be a
possible way to better understand the correlation structure among
attributes, and this could also make the algorithm more efficient.

Third, this research is based on the DINA model. However because of
the conjunctive nature of the model that divides examinees only into
either the mastery or non-mastery category, further research might
apply the estimation procedure to more general models, such as the
G-DINA model, which can identify the probability of different attribute
patterns.

As for the measure of accuracy, researchers might argue that in calculating
the recovery rate $\Delta_{q}$, $\hat{\boldsymbol{q}}$ should be
rounded to the nearest whole before subtracting the actual Q-matrix.
That is, instead of using equation (7), the recovery rate should be
defined as \begin{equation}
\Delta_{q}=\frac{1}{M}\sum\limits _{m=1}^{M}\biggl(1-\frac{\Bigl|\bigl[\hat{\boldsymbol{q}}^{(m)}\bigr]-\boldsymbol{q}\Bigr|}{JK}\biggr),\:m=1,2,\ldots,M, \tag{8}
\end{equation} 

\noindent where the $\left[\cdot\right]$ returns the value rounded
to the nearest whole. This concern matters only when Q-matrix estimates
are mostly close to 0.5. As a matter of fact, when tested, using equation
(8) increased the recovery rate in each of the simulation studies. 

Another issue concerns the software. Estimating the Q-matrix is computationally
intensive. Although the customized R program ran well, it took about
26 hours for a run of MCMC in the simulation. Therefore it would be
worth the effort to convert the code to another lower-level programming
language, such as C or Java, to facilitate efficiency.

Among the many issues, how to interpret the estimated Q-matrix might
be the most challenging. Although our Q-matrix estimate for the ECPE
data is somewhat close to the initial Q-matrix in Table 3, we are
not sure whether these 5 attributes derived from the data correspond
to those 3 attributes appeared in Henson and Templin (2007). Based
on the AIC, the preferred Q-matrix is the one estimated Q-matrix.
Nevertheless, we certainly do not claim our Q-matrix estimate is the
correct answer. This estimated Q-matrix should be treated circumspectly.
Discussion of the meaning of each entry is beyond the scope of this
paper, and the interpretation and implication are left to domain experts.
\newpage{}
\noindent \begin{center}
\textbf{\Large{}References}\vspace{-30pt}
\par\end{center}
\begin{description}
\begin{singlespace}
\item [{\textmd{Barnes,}}] T. M. (2003). \textit{The Q-matrix Method of
Fault-tolerant Teaching in Knowledge Assessment and Data Mining} (Doctoral
Dissertation). North Carolina State University.
\item [{\textmd{Chen,}}] Y., Liu, J., Xu, G., \& Ying, Z. (2015). Statistical
analysis of Q-matrix based diagnostic classification models. \textit{Journal
of the American Statistical Association}, 110(510), 850-866.
\item [{\textmd{Chiu,}}] C. Y. (2013). Statistical Refinement of the Q-matrix
in Cognitive Diagnosis. \textit{Applied Psychological Measurement},
37(8), 598-618.
\item [{\textmd{Chiu,}}] C.-Y., Douglas J., \& Li, X. (2009). Cluster analysis
for cognitive diagnosis: Theory and applications. \textit{Psychometrika},
74, 633-665.
\item [{\textmd{Chung,}}] M. (2014). Estimating the Q-matrix for cognitive
diagnosis models in a Bayesian framework, Unpublished doctoral thesis,
Columbia University
\item [{\textmd{Chung,}}] M., \& Johnson, M. S. (2017). \textit{Developing
an MCMC Algorithm for the Estimation of the Bayesian Reduced RUM}.
Manuscript submitted for publication.
\item [{\textmd{Congdon,}}] P. (2005), \textit{Bayesian Models for Categorical
Data}., John Wiley \& Sons, Ltd, Chichester, UK.
\item [{\textmd{DeCarlo,}}] L. T. (2011). On the analysis of fraction subtraction
data: The DINA model, classification, latent class sizes, and the
Q-matrix. \textit{Applied Psychological Measurement}, 35, 8-26.
\item [{\textmd{DeCarlo,}}] L. T. (2012). Recognizing uncertainty in the
Q-matrix via a Bayesian extension of the DINA model. \textit{Applied
Psychological Measurement}, 36, 447-468.
\item [{\textmd{de}}] la Torre, J. (2008). An empirically-based method
of Q-matrix validation for the DINA model: Development and applications.
\textit{Journal of Educational Measurement}, 45, 343-362.
\item [{\textmd{de}}] la Torre, J. (2009). DINA model and parameter estimation:
A didactic. \textit{Journal of Educational and Behavioral Statistics},
34, 115-130.
\item [{\textmd{de}}] la Torre, J., \& Douglas, J. (2004). Higher-order
latent trait models for cognitive diagnosis. \textit{Psychometrika},
69:333-353.
\item [{\textmd{Desmarais,}}] M. C. (2012). Mapping question items to skills
with non-negative matrix factorization. \textit{ACM SIGKDD Explorations
Newsletter}, 13(2), 30-36.
\item [{\textmd{DiBello,}}] L. V., Stout, W. F., \& Roussos, L. A. (1995).
Unified cognitive psychometric assessment likelihood-based classification
techniques, chapter Cognitively diagnostic assessment, pages 361-390.
Hillsdale, NJ: Erlbaum.
\item [{\textmd{ECPE}}] (2015). ECPE 2015 Report (p. 1). The Examination
for the Certificate of Proficiency in English (ECPE).
\item [{\textmd{Erosheva,}}] E. A., \& Curtis, S. M. (2017). Dealing with
Reflection Invariance in Bayesian Factor Analysis. \textit{Psychometrika},
1-13.
\item [{\textmd{Feng,}}] Y., Habing, B. T., \& Huebner, A. (2014). Parameter
estimation of the Reduced RUM using the EM algorithm.\textit{ Applied
Psychological Measurement}, 38, 137\textendash 150.
\item [{\textmd{Gelman,}}] A., \& Shirley, K. (2011). Inference and Monitoring
Convergence. In Steve Brooks, A. Gelman, G. L. Jones, \& X.-L. Meng
(eds.), \textit{Handbook of Markov Chain Monte Carlo}, pp. 163-174,
Chapman \& Hall/CRC, New York, USA.
\item [{\textmd{Geyer,}}] C. J. (1991). Markov Chain Monte Carlo maximum
likelihood, in \textit{Computing Science and Statistics: Proceedings
of the 23rd Symposium on the Interface}, edited by E. M. Keramidas,
pp. 156\textendash 163, Interface Found., Fairfax Station, Va., 1991.
\item [{\textmd{Hartz,}}] S. (2002). \textit{A Bayesian framework for the
Unified Model for assessing cognitive abilities: Blending theory with
practicality} (Doctoral dissertation). University of Illinois, Urbana-Champaign.
\item [{\textmd{Henson,}}] R., \& Templin, J. (2007, April). \textit{Importance
of Q-matrix construction and its effects cognitive diagnosis model
results}. Paper presented at the annual meeting of the National Council
on Measurement in Education in Chicago, Illinois.
\item [{\textmd{Henson,}}] R. A., Templin, J. L., \& Willse, J. T. (2009).
Defining a Family of Cognitive Diagnosis Models Using Log-Linear Models
with Latent Variables. \textit{Psychometrika}, 74(2):191-210.
\item [{\textmd{Jasra,}}] A., Holmes, C. C., \& Stephens D. A. (2005).
Markov Chain Monte Carlo methods and the label switching problem in
Bayesian mixture modeling. \textit{Statistical Science}, 20, 50\textendash 67.
\item [{\textmd{Junker,}}] B. W., \& Sijtsma, K. (2001). Cognitive assessment
models with few assumptions, and connections with nonparametric item
response theory. \textit{Applied Psychological Measurement}, 25, 258-272.
\item [{\textmd{Leighton,}}] J. P., \& Gierl, M. J. (Eds.). (2007). \textit{Cognitive
diagnostic assessment for education. Theory and applications}. Cambridge,
MA: Cambridge University Press.
\item [{\textmd{Liu,}}] J., Xu, G., \& Ying, Z. (2012). Data-driven learning
of Q-matrix. \textit{Applied Psychological Measurement}, 36, 609-618.
\item [{\textmd{Maris,}}] E. (1999). Estimating multiple classification
latent class models. \textit{Psychometrika}, 64, 187\textendash 212.
\item [{\textmd{R}}] Development Core Team. (2017). R: A language and environment
for statistical computing {[}Computer software{]}. Vienna, Austria:
R Foundation for Statistical Computing. Available from http://www.r-project.org.
\item [{\textmd{Ross,}}] S. M. (2006), \textit{Simulation}. 4th ed., Academic
Press, San Diego.
\item [{\textmd{Tatsuoka,}}] C. (2002). Data analytic methods for latent
partially ordered classification models. \textit{Journal of the Royal
Statistical Society, Series C, Applied Statistics}, 51, 337\textendash 350.
\item [{\textmd{Tatsuoka,}}] K. K. (1983). Rule space: An approach for
dealing with misconceptions based on item response theory. \textit{Journal
of Educational Measurement}, 20, 345\textendash 354.
\item [{\textmd{Tatsuoka,}}] K. K. (1990). Toward an integration of item-response
theory and cognitive error diagnosis. In N. Frederiksen, R. Glaser,
A. Lesgold, \& M. Shafto (Eds.), \textit{Diagnostic monitoring of
skill and knowledge acquisition} (pp. 453-488). Hillsdale, NJ: Erlbaum.
\item [{\textmd{Templin,}}] J., \& Bradshaw, L. (2014). Hierarchical diagnostic
classification models: A family of models for estimating and testing
attribute hierarchies. \textit{Psychometrika}, 79, 317-339.
\item [{\textmd{Templin,}}] J., \& Henson, R. (2006, April). \textit{A
Bayesian method for incorporating uncertainty into Q-matrix estimation
in skills assessment}. Paper presented at the annual meeting of the
National Council on Measurement in Education, San Francisco, CA.
\item [{\textmd{Templin,}}] J., \& Hoffman, L. (2013). Obtaining diagnostic
classification model estimates using Mplus. \textit{Educational Measurement:
Issues and Practice}, 32, 37-50.
\item [{\textmd{Winters,}}] T. (2006).\textit{ Educational Data Mining:
Collection and Analysis of Score Matrices for Outcomes-Based Assessment}
(Doctoral dissertation). University of California, Riverside.
\item [{\textmd{Xu,}}] P., Desmarais, M. C. (2016). Boosted decision tree
for Q-matrix refinement. In: 9th International Conference on Educational
Data Mining, 6 June\textendash 2 July 2016, Raleigh, NC, USA.
\end{singlespace}
\end{description}

\end{document}